\documentclass[aps,prd,superscriptaddress,nofootinbib,amsmath,amsfonts,preprintnumbers,groupedaddress,showpacs,9pt,english]{revtex4}
\usepackage{amsmath}
\usepackage{amssymb}
\usepackage{babel}
\usepackage{wrapfig}
\usepackage{cancel}
\usepackage{dblfloatfix}
\usepackage{float}
\usepackage[none]{hyphenat}
\usepackage{placeins}

\usepackage{relsize,exscale}
\makeatletter

\newcommand{\beq}{\begin{equation}}
\newcommand{\eeq}{\end{equation}}
\newcommand{\bea}{\begin{eqnarray}}
\newcommand{\eea}{\end{eqnarray}}

\newcommand{\comm}[1]{}

\usepackage{array,multirow,graphicx}
\usepackage{dcolumn}
\usepackage{newlfont}
\usepackage{bm}
\usepackage[colorlinks,citecolor=blue,urlcolor=blue,linkcolor=blue]{hyperref}
\usepackage[figtopcap]{subfigure}
\usepackage{color}

\usepackage{scalerel}
\usepackage{tikz}
\usetikzlibrary{svg.path}
\definecolor{orcidlogocol}{HTML}{A6CE39}
\tikzset{
  orcidlogo/.pic={
    \fill[orcidlogocol] svg{M256,128c0,70.7-57.3,128-128,128C57.3,256,0,198.7,0,128C0,57.3,57.3,0,128,0C198.7,0,256,57.3,256,128z};
    \fill[white] svg{M86.3,186.2H70.9V79.1h15.4v48.4V186.2z}
                 svg{M108.9,79.1h41.6c39.6,0,57,28.3,57,53.6c0,27.5-21.5,53.6-56.8,53.6h-41.8V79.1z M124.3,172.4h24.5c34.9,0,42.9-26.5,42.9-39.7c0-21.5-13.7-39.7-43.7-39.7h-23.7V172.4z}
                 svg{M88.7,56.8c0,5.5-4.5,10.1-10.1,10.1c-5.6,0-10.1-4.6-10.1-10.1c0-5.6,4.5-10.1,10.1-10.1C84.2,46.7,88.7,51.3,88.7,56.8z};}}
\newcommand\orcid[1]{\href{https://orcid.org/#1}{\mbox{\scalerel*{
\begin{tikzpicture}[yscale=-1,transform shape]
\pic{orcidlogo};
\end{tikzpicture}
}{|}}}}
\begin{document}

\date{\today}

\title{\textbf{Photon Sphere and Shadow of a Perturbative Black Hole in $f(R,\mathcal{G})$ Gravity}}

\
\author{G.G.L. Nashed}\email{nashed@bue.edu.eg}
\affiliation{Centre for Theoretical Physics, The British University, P.O. Box 43, El Sherouk City, Cairo 11837, Egypt\\ Center for Space Research, North-West University, Potchefstroom 2520, South Africa}

\begin{abstract}
We investigate the impact of higher-curvature corrections on black-hole observables within a perturbative $f(R, G)$ gravity framework. Working in a static, spherically symmetric spacetime, we construct leading-order deviations from the Schwarzschild solution by expanding the field equations in small coupling parameters associated with quadratic curvature invariants. The resulting metric corrections are obtained as asymptotic expansions and used to analyze null geodesics. We derive analytic expressions for the shift in the photon-sphere radius and show that higher-curvature terms modify the location of unstable photon orbits, with the Gauss--Bonnet sector producing a more significant contribution than mixed curvature terms. These modifications propagate to observable quantities, leading to corrections in the black-hole shadow radius. We identify the distinct roles of photon-sphere displacement and direct metric perturbations in determining the shadow size. We further discuss the implications of these corrections for strong gravitational lensing and quasinormal modes, highlighting the enhanced sensitivity of strong-field observables to higher-curvature effects. While the present analysis is based on an asymptotic perturbative treatment, our results provide a consistent framework for estimating leading-order deviations from general relativity and suggest that high-resolution observations, including very-long-baseline interferometry and gravitational-wave measurements, may offer constraints on modified gravity models.
\end{abstract}
\maketitle
\section{Introduction}
General Relativity (GR) has been remarkably successful in describing gravitational phenomena across a wide range of scales, from solar-system tests to the dynamics of compact astrophysical objects. However, several theoretical and observational motivations suggest that GR may require modifications in the strong-curvature regime or at high energies. These include the quest for a quantum theory of gravity, the resolution of cosmological puzzles such as dark energy and inflation, and the possibility of deviations from GR in the vicinity of black holes \cite{Clifton:2011jh,Nojiri:2010wj,Capozziello:2011et}.

{ Among the most widely studied extensions of general relativity are higher-curvature theories, in which the Einstein--Hilbert action is generalized through nonlinear functions of curvature invariants. Modified gravitational theories coupled to nonlinear matter sectors have also attracted considerable attention in recent years. In particular, the gravitational signatures of nonlinear electrodynamics in the framework of \(f(R,T)\) gravity were investigated in Ref.~\cite{AraujoFilho:2025hnf}, where it was shown that nonminimal matter--geometry couplings can significantly affect the spacetime structure and black-hole observables. These results further highlight the sensitivity of strong-field gravitational phenomena to corrections beyond general relativity.

In this context, modified Gauss--Bonnet gravity has been extensively studied as a viable extension of Einstein gravity. Models in which an arbitrary function of the Gauss--Bonnet invariant, \(f(\mathcal{G})\), is added to the Einstein--Hilbert action were shown to lead to modified metric field equations capable of naturally explaining the late-time accelerated expansion of the universe without the introduction of an explicit dark-energy component~\cite{Cognola:2006sp}. The cosmological implications and theoretical viability of such higher-curvature models were also investigated in connection with the hierarchy problem. Comprehensive analyses of modified gravity theories, including \(f(R)\), \(f(\mathcal{G})\), and more general higher-curvature extensions, were later presented in Refs.~\cite{Nojiri:2006ri,Nojiri:2010wj}, where the metric formulation and cosmological consequences of generalized gravitational actions were systematically reviewed.

A reconstruction of the \(\Lambda\)CDM cosmological epoch within the framework of \(F(R,\mathcal{G})\) and modified Gauss--Bonnet gravity was presented in Ref.~\cite{Elizalde:2010ts}. It was demonstrated that generalized higher-curvature models can successfully reproduce the observed accelerated expansion of the universe without explicitly introducing a cosmological constant. The cosmological viability of these theories and their capability to describe different phases of cosmic evolution were also investigated.

Cosmological perturbations in \(f(R,\mathcal{G})\) gravity coupled to a perfect fluid were analyzed in Ref.~\cite{DeFelice:2010aj}. Scalar, vector, and tensor perturbation modes were examined, and important conditions required for the absence of ghost instabilities and other pathological behaviors were derived. These results provided significant theoretical constraints on the consistency and viability of generalized \(f(R,\mathcal{G})\) gravitational theories.

Further investigations explored both the cosmological dynamics and theoretical consistency of modified Gauss--Bonnet gravity. The cosmological evolution induced by \(f(\mathcal{G})\) corrections and their influence on the expansion history of the universe were studied in Ref.~\cite{Li:2007jm}. Cosmologically viable \(f(\mathcal{G})\) dark-energy models satisfying observational and stability requirements were later constructed in Ref.~\cite{DeFelice:2008wz}. In addition, the stability of Schwarzschild-like solutions in \(f(R,\mathcal{G})\) gravity was investigated in Ref.~\cite{DeFelice:2009ak}, providing important constraints on the theoretical viability of generalized higher-curvature gravitational theories.}In particular, $f(R,\mathcal{G})$ gravity, where $\mathcal{G}$ denotes the Gauss--Bonnet invariant,
\begin{equation}\label{Guss}
\mathcal{G} = R^2 - 4R_{\mu\nu}R^{\mu\nu} + R_{\mu\nu\rho\sigma}R^{\mu\nu\rho\sigma},
\end{equation}
has attracted considerable attention. This class of theories naturally arises in low-energy limits of string theory and provides a rich phenomenology while avoiding some of the pathologies associated with more general higher-derivative models \cite{Nojiri:2005vv,DeFelice:2010aj}.

Black holes offer a unique arena for testing modified theories of gravity in the strong-field regime. In particular, the properties of null geodesics around black holes encode key observational signatures, such as the photon sphere and the black-hole shadow. The photon sphere corresponds to unstable circular photon orbits and plays a central role in determining both gravitational lensing and the apparent size of the shadow \cite{Claudel:2000yi,Perlick:2004tq}. The shadow itself, defined as the dark region in the observer's sky corresponding to photon capture, has recently become an observable quantity thanks to the Event Horizon Telescope (EHT), which has produced horizon-scale images of the supermassive black holes in M87* and Sgr A* \cite{EventHorizonTelescope:2019dse,EventHorizonTelescope:2022wkp}. These effects may, in principle, be constrained by high-resolution observations of black-hole shadows, such as those provided by the Event Horizon Telescope.

The size and shape of the black-hole shadow are sensitive to the underlying spacetime geometry and therefore provide a powerful probe of deviations from GR \cite{Broderick:2013rlq,EventHorizonTelescope:2020qrl}. In modified gravity theories, corrections to the metric functions can shift the photon-sphere radius and consequently alter the shadow size. These effects can, in principle, be constrained by observations, making shadow studies an important tool for testing alternative theories of gravity. Recent studies have also investigated quantum-corrected Schwarzschild geometries within effective-field-theory approaches to gravity. In particular, Battista~\cite{Battista:2023iyu} analyzed low-energy one-loop quantum corrections to the Schwarzschild spacetime and showed that such corrections can modify the behavior of null and timelike geodesics, light deflection, and strong-field optical observables while preserving a perturbative description of the geometry. Black-hole shadows in modified gravity theories have recently been investigated in detail within the framework of $F(R)$ gravity. In particular, Nojiri and Odintsov~\cite{Nojiri:2024qgx} analyzed the influence of modified curvature terms on the photon-sphere structure and shadow observables, showing that deviations from general relativity can produce measurable corrections to the shadow radius and optical appearance of black holes. Their results further support the use of shadow observations as probes of higher-curvature gravitational effects.


More recently, Wang and Battista~\cite{Wang:2025fmz} investigated the dynamical properties and optical appearance of quantum-corrected Schwarzschild black holes, including the behavior of stable and unstable circular orbits, black-hole shadows and rings, as well as thermodynamic quantities such as the Hawking temperature and entropy. In particular, logarithmic corrections to the black-hole entropy were identified, highlighting the interplay between quantum effects and strong-gravity observables. These studies further support the idea that photon-sphere observables and black-hole imaging may provide important probes of quantum corrections to gravitational dynamics.

Recent studies have further demonstrated that black-hole shadow observations can provide important constraints on extensions of general relativity involving higher-dimensional and higher-curvature effects. In particular, Banerjee \textit{et al.}~\cite{Banerjee:2019nnj} analyzed the shadow of M87* in the presence of hidden extra dimensions and showed that modifications to the photon-sphere structure can leave potentially observable imprints on the shadow radius and geometry. Their results support the idea that horizon-scale observations may serve as sensitive probes of beyond-GR physics. 

In this work, we investigate static, spherically symmetric black-hole solutions in a perturbative $f(R,\mathcal{G})$ framework. By expanding the field equations around the Schwarzschild solution, we obtain leading-order asymptotic analytic expressions for the corrections to the metric functions induced by higher-curvature terms. We then analyze the impact of these corrections on null geodesics, focusing in particular on the photon sphere and the black-hole shadow.

Our results show that the higher-curvature couplings introduce shifts in the photon-sphere radius and the shadow size, with distinct contributions arising from different curvature terms. In particular, we find that the Gauss--Bonnet sector can have a more pronounced effect on observable quantities than mixed curvature terms, highlighting the importance of considering multiple invariants in modified gravity models.

The structure of this paper is as follows. Section~\ref{II} outlines the field equations together with the perturbative approach employed throughout the analysis. In Section~\ref{III}, we specify the metric ansatz and compute the corrections at leading order. The properties of null geodesics are investigated in Section~\ref{IV}, where the photon-sphere equation and its perturbative modification are derived. Section~\ref{V} is devoted to the analysis of the black-hole shadow, while Section~\ref{VI} examines the strong gravitational lensing effects. In Section~\ref{VII}, we explore the consequences for quasinormal modes and the associated ringdown behavior. Finally, Section~\ref{VIII} contains a summary of the main results along with remarks on possible future developments.

\section{$f(R,\mathcal{G})$ Gravity}\label{II}

A direct extension of Einstein's theory of gravity can be formulated by allowing the gravitational Lagrangian to incorporate higher--order curvature invariants. Among these modified theories, $f(R,\mathcal{G})$ gravity has attracted significant interest, where $R$ denotes the Ricci scalar and $\mathcal{G}$ represents the Gauss--Bonnet invariant, defined in Eq.~\eqref{Guss}.  The action of the theory is given by
\begin{equation}
S = \frac{1}{16\pi} \int d^4x \, \sqrt{-g} \, f(R,\mathcal{G}).
\end{equation}
In four spacetime dimensions, the Gauss--Bonnet term by itself is a topological invariant and does not contribute to the field equations. However, when it appears inside a nontrivial function $f(R,\mathcal{G})$, it contributes dynamically and modifies the gravitational equations.

To derive the field equations, we vary the action with respect to the metric $g_{\mu\nu}$. The variation reads
\begin{equation}
\delta S = \frac{1}{16\pi} \int d^4x \left[ \delta(\sqrt{-g}) f + \sqrt{-g} \, \delta f \right].
\end{equation}
Using the identity
\begin{equation}
\delta \sqrt{-g} = -\frac{1}{2} \sqrt{-g} \, g_{\mu\nu} \delta g^{\mu\nu},\quad \mbox{we obtain}\quad
\delta S = \frac{1}{16\pi} \int d^4x \, \sqrt{-g} \left[ -\frac{1}{2} g_{\mu\nu} f \, \delta g^{\mu\nu} + \delta f \right].
\end{equation}
The variation of the function $f(R,\mathcal{G})$ is given by
\begin{equation}
\delta f = f_R \, \delta R + f_{\mathcal{G}} \, \delta \mathcal{G},\quad \mbox{where} \quad
f_R \equiv \frac{\partial f}{\partial R}, \qquad
f_{\mathcal{G}} \equiv \frac{\partial f}{\partial \mathcal{G}}. 
\end{equation}
The variation of the Ricci scalar is
\begin{equation}
\delta R = R_{\mu\nu} \delta g^{\mu\nu} + \nabla_\mu \left( \nabla_\nu \delta g^{\mu\nu} - g_{\alpha\beta} \nabla^\mu \delta g^{\alpha\beta} \right).
\end{equation}
After integrating by parts and discarding boundary terms, this leads to
\begin{align}
&\delta (\sqrt{-g} f_R R) \rightarrow \sqrt{-g} \left[ f_R R_{\mu\nu} - \nabla_\mu \nabla_\nu f_R + g_{\mu\nu} \Box f_R \right] \delta g^{\mu\nu},\quad \mbox{where the d'Alembertian operator is defined as} \nonumber\\
 &\Box \equiv g^{\mu\nu} \nabla_\mu \nabla_\nu.
\end{align}
{ The variation of the Gauss--Bonnet sector gives the contribution
\begin{align}
H_{\mu\nu} ={}&
2R g_{\mu\nu}\Box f_{\mathcal{G}}
-2R\nabla_\mu\nabla_\nu f_{\mathcal{G}}
-4R_{\mu\nu}\Box f_{\mathcal{G}}
-4g_{\mu\nu}R^{\alpha\beta}\nabla_\alpha\nabla_\beta f_{\mathcal{G}}
\nonumber\\
&+4R_{\mu}^{\ \alpha}\nabla_\nu\nabla_\alpha f_{\mathcal{G}}
+4R_{\nu}^{\ \alpha}\nabla_\mu\nabla_\alpha f_{\mathcal{G}}
+4R_{\mu\alpha\nu\beta}\nabla^\alpha\nabla^\beta f_{\mathcal{G}}
\nonumber\\
&+2f_{\mathcal{G}}RR_{\mu\nu}
-4f_{\mathcal{G}}R_{\mu}^{\ \alpha}R_{\alpha\nu}
-4f_{\mathcal{G}}R_{\mu\alpha\nu\beta}R^{\alpha\beta}
+2f_{\mathcal{G}}R_{\mu}^{\ \alpha\beta\gamma}
R_{\nu\alpha\beta\gamma}.
\end{align}
Collecting all contributions, the metric field equations of \(f(R,\mathcal{G})\) gravity can be written as~\cite{Elizalde:2010ts,DeFelice:2010aj}
\begin{equation}\label{fe}
f_R R_{\mu\nu}
-\frac{1}{2}g_{\mu\nu}f
+\left(g_{\mu\nu}\Box-\nabla_\mu\nabla_\nu\right)f_R
+H_{\mu\nu}=0.
\end{equation}}

\subsection*{Perturbative Model}
We consider the perturbative model
\begin{equation}\label{mod}
f(R,\mathcal{G}) = R + \alpha R^2 + \beta R \mathcal{G} + \gamma \mathcal{G}^2,\quad \mbox{where $\alpha$, $\beta$, and $\gamma$ are small \emph{dimensional} coupling constants.}
\end{equation}
Since the Ricci scalar scales as $[R] = L^{-2}$ and the Gauss--Bonnet invariant as $[\mathcal{G}] = L^{-4}$, the dimensions of the coupling constants are
\begin{equation}
[\alpha] = L^{2}, \qquad [\beta] = L^{6}, \qquad [\gamma] = L^{8}.
\end{equation}
Using Eq.~\eqref{mod}, the derivatives of the function are given by
\begin{equation}\label{diff}
f_R = 1 + 2\alpha R + \beta \mathcal{G}, \qquad
f_{\mathcal{G}} = \beta R + 2\gamma \mathcal{G}.
\end{equation}

Substituting into the field equations, one obtains modified Einstein equations of the schematic form
\begin{equation}
G_{\mu\nu} + \alpha E_{\mu\nu}^{(R^2)} + \beta E_{\mu\nu}^{(R\mathcal{G})} + \gamma E_{\mu\nu}^{(\mathcal{G}^2)} = 0,
\end{equation}
where each term represents the contribution from the corresponding higher-curvature invariant. It is important to note that, on the Schwarzschild background, the Ricci scalar and Ricci tensor vanish identically, $R^{(0)} = 0$ and $R_{\mu\nu}^{(0)} = 0$. As a consequence, the quadratic Ricci scalar term $\alpha R^2$ does not contribute to the leading-order perturbative corrections, since both $R^{(0)}$ and its derivatives vanish at this order. In contrast, the Gauss--Bonnet invariant remains nonzero, $G^{(0)} \propto M^2 / r^6$, and therefore the $\beta$ and $\gamma$ sectors generate nontrivial contributions to the field equations.

Accordingly, within the present perturbative framework around the Schwarzschild solution, the leading deviations from general relativity are governed by the $RG$ and $G^2$ terms, while the $\alpha$-sector enters only at higher order or in backgrounds with nonvanishing Ricci curvature.

The perturbative treatment is valid provided that the higher-curvature corrections remain small compared to the leading Einstein term. In practice, this requires that the dimensionless combinations constructed from the coupling constants and the curvature invariants satisfy
\begin{equation}
|\alpha R| \ll 1, \qquad |\beta G| \ll 1, \qquad |\gamma G| \ll 1,
\end{equation}
in the region of interest. For the Schwarzschild background, where $R^{(0)}=0$ and $G^{(0)} \sim M^2/r^6$, these conditions translate into constraints on the ratios of the coupling constants to appropriate powers of the mass scale $M$, ensuring that the perturbative expansion remains well controlled.

In this regime, all quantities can be expanded to first order in $(\alpha, \beta, \gamma)$, allowing a systematic analysis of deviations from general relativity while maintaining analytical control.
\section{Metric Ansatz}\label{III}

To investigate static and spherically symmetric solutions in modified gravity, we consider the most general form of a four-dimensional spacetime invariant under time translations and spatial rotations. In Schwarzschild-like coordinates, the line element can be written as
\begin{align}\label{met}
&ds^2 = -A(r) dt^2 + B(r) dr^2 + r^2 d\Omega^2,\quad \mbox{where $A(r)$ and $B(r)$ are functions of the radial coordinate $r$, and}\nonumber\\
&\quad d\Omega^2 = d\theta^2 + \sin^2\theta, d\phi^2, \quad \mbox{is the metric on the unit two-sphere.}
\end{align}
 This form follows uniquely from the imposed symmetries and leaves two independent functions in the $(t,r)$ sector.

In general relativity, Birkhoff's theorem guarantees that the unique vacuum, static, spherically symmetric solution is the Schwarzschild metric,
\begin{equation}
A_0(r) = 1 - \frac{2M}{r}, \qquad
B_0(r) = \left(1 - \frac{2M}{r}\right)^{-1}, \qquad \mbox{where $M$ is the gravitational mass.}
\end{equation}
 In contrast, in modified gravity theories such as $f(R,\mathcal{G})$ gravity, Birkhoff's theorem does not generally hold \cite{DeFelice:2010aj,Nojiri:2005jg}, and additional degrees of freedom can modify the vacuum geometry. Consequently, deviations from the Schwarzschild solution are expected even in the absence of matter sources.

To capture these deviations in a controlled manner, we adopt a perturbative approach and expand the metric functions around the Schwarzschild background as
\begin{equation}\label{prt}
A(r) = A_0(r) + \epsilon a(r), \qquad
B(r) = B_0(r) + \epsilon b(r),\quad \mbox{where $\epsilon \ll 1$ is a dimensionless bookkeeping parameter}.
\end{equation}
Here $a(r)$ and $b(r)$ represent the leading-order corrections induced by higher-curvature effects. At zeroth order in $\epsilon$, one recovers the Schwarzschild solution, while at first order the modified field equations reduce to linear differential equations governing the perturbations. It is important to emphasize that $\epsilon$ is introduced solely to organize the perturbative expansion and does not represent a physical parameter. In practice, one expands all quantities to first order in the small couplings $(\alpha,\beta,\gamma)$. The perturbations $a(r)$ and $b(r)$ are not entirely independent due to coordinate (gauge) freedom. In particular, a radial redefinition of the form
\begin{equation}
r \rightarrow r + \epsilon \xi(r)
\end{equation}
induces shifts in these functions. In the present analysis, we work in Schwarzschild-like coordinates and keep both functions explicit, with the understanding that physical observables must be gauge-invariant.

The function $A(r)$ determines the redshift factor and the effective gravitational potential, while $B(r)$ governs the radial proper distance and contributes to the radial motion of test particles. As a result, modifications to these functions directly affect null geodesics and therefore influence the photon-sphere radius and the black-hole shadow.

For physically relevant solutions, appropriate boundary conditions must be imposed. Asymptotic flatness requires
\begin{equation}
\lim_{r \to \infty} A(r) = 1, \qquad
\lim_{r \to \infty} B(r) = 1, \quad \mbox{which implies}\quad
\lim_{r \to \infty} a(r) = 0, \qquad
\lim_{r \to \infty} b(r) = 0.
\end{equation}
Near the Schwarzschild horizon at $r = 2M$, regularity requires that the perturbations do not introduce divergences stronger than those already present in the background metric.

Substituting this ansatz into the modified field equations of $f(R,\mathcal{G})$ gravity and expanding to first order in $\epsilon$ yields a coupled system of ordinary differential equations for $a(r)$ and $b(r)$. Solving this system, subject to the boundary conditions specified above, determines the leading deviations from the Schwarzschild geometry and provides the necessary input for the analysis of photon trajectories in the following section. The main steps and source structure are summarized in Appendix A.

Here, we present the final expressions for the metric functions, which take the form
\begin{equation}
A(r) = 1 - \frac{2M}{r} + \epsilon\, a(r),
\qquad
B(r) = \left(1 - \frac{2M}{r}\right)^{-1} + \epsilon\, b(r),
\end{equation}
where the functions $a(r)$ and $b(r)$ are given by:
\begin{align}
a(r) &=
\beta\left(
-\frac{10368}{49}\frac{M^2}{r^7}
+\frac{46359}{98}\frac{M^3}{r^8}
+\frac{15067}{441}\frac{M^4}{r^9}
+\frac{4309}{75}\frac{M^5}{r^{10}}
+\cdots
\right)
\nonumber\\
&\quad
+\gamma\left(
-\frac{102528}{25}\frac{M^3}{r^{10}}
+\frac{24772608}{3025}\frac{M^4}{r^{11}}
+\frac{684864}{3025}\frac{M^5}{r^{12}}
+\frac{209725056}{511225}\frac{M^6}{r^{13}}
+\cdots
\right),
\\[1ex]
b(r) &=
\beta\left(
\frac{8352}{49}\frac{M^2}{r^7}
-\frac{19359}{98}\frac{M^3}{r^8}
-\frac{5905}{441}\frac{M^4}{r^9}
-\frac{325993}{11025}\frac{M^5}{r^{10}}
+\cdots
\right)
\nonumber\\
&\quad
+\gamma\left(
-\frac{1152}{25}\frac{M^3}{r^{10}}
+\frac{4068864}{3025}\frac{M^4}{r^{11}}
-\frac{46656}{3025}\frac{M^5}{r^{12}}
-\frac{5924736}{102245}\frac{M^6}{r^{13}}
+\cdots
\right).
\end{align}
Only the leading terms relevant for the present asymptotic analysis are displayed.
\section{Photon Sphere}\label{IV}

The study of null geodesics in strong gravitational fields provides direct insight into observable phenomena such as black hole shadows and gravitational lensing. In particular, static, spherically symmetric spacetimes admit circular photon orbits, known as the \emph{photon sphere}, which play a central role in determining the optical appearance of black holes \cite{Synge:1966okc,Chandrasekhar:1984siy,Claudel:2000yi}.

A general static and spherically symmetric spacetime can be described by the metric given by Eq.~\eqref{met} the motion of photons follows null geodesics, $ds^2=0$. Due to the symmetries of the spacetime, the energy $E$ and angular momentum $L$ of a test particle are conserved quantities. Using these conserved quantities, the radial equation for null geodesics can be expressed as \cite{Chandrasekhar:1984siy}
\begin{equation}
\dot{r}^2 = \frac{1}{B(r)}\left[\frac{E^2}{A(r)} - \frac{L^2}{r^2}\right],
\end{equation}
where the dot denotes differentiation with respect to an affine parameter.

The right-hand side can be interpreted as an effective potential for radial motion. Circular photon orbits occur when the radial velocity vanishes and the effective potential is extremized, i.e.
\begin{equation}
\dot{r} = 0, \qquad \frac{d}{dr}(\dot{r}^2) = 0.
\end{equation}
These conditions imply that the function $r^2/A(r)$ must be extremized, leading to the photon-sphere condition \cite{Claudel:2000yi,Perlick:2004tq},
 \begin{equation}
\frac{d}{dr}\left(\frac{r^2}{A(r)}\right) = 0.
\end{equation}
Evaluating the derivative explicitly, one finds
\begin{equation}
\frac{d}{dr}\left(\frac{r^2}{A(r)}\right)
= \frac{2rA(r) - r^2 A'(r)}{A(r)^2}.
\end{equation}
Since the denominator is nonzero, the condition reduces to
\begin{equation}\label{ps}
r A'(r) - 2A(r) = 0.
\end{equation}
This relation depends solely on the temporal component of the metric and is a standard result for static, spherically symmetric spacetimes \cite{Perlick:2004tq}. The behavior of photon spheres has also been investigated in dynamically evolving spacetimes. In this context, Mishra \textit{et al.}~\cite{Mishra:2019trb} demonstrated that the evolution of the photon sphere can significantly affect the shadow structure and null geodesic behavior in non-static geometries. Although the present work focuses on static and spherically symmetric configurations, the analysis presented here may provide a useful starting point for future investigations of dynamical black-hole solutions in perturbative $f(R,\mathcal{G})$ gravity.

In general relativity, the Schwarzschild solution is given by
\begin{equation}
A_0(r) = 1 - \frac{2M}{r}.
\end{equation}
Substituting into the photon-sphere condition yields
\begin{equation}
r \frac{2M}{r^2} - 2\left(1 - \frac{2M}{r}\right) = 0,\qquad \mbox{which yields the well-known result} \qquad
r_{\mathrm{ph}}^{(0)} = 3M.
\end{equation}
This is the well-known location of the unstable circular photon orbit in Schwarzschild spacetime \cite{Chandrasekhar:1984siy}.

We now consider deviations from the Schwarzschild geometry induced by higher-curvature corrections. Using the perturbative expansion introduced in the previous section,
\begin{equation}
A(r) = A_0(r) + \epsilon a(r),\quad \mbox{we assume that the photon-sphere radius is shifted as} \quad
r_{\mathrm{ph}} = 3M + \epsilon\delta r,
\end{equation}
where $\delta r$ represents the first-order correction. Substituting into the photon-sphere condition given by Eq.~\eqref{ps},
\begin{equation}
r \big[A_0'(r) + \epsilon a'(r)\big] - 2 \big[A_0(r) + \epsilon a(r)\big] = 0, \quad \mbox{which gives} \quad
\big[rA_0'(r)-2A_0(r)\big]+\epsilon\big[ra'(r)-2a(r)\big]=0.
\end{equation}

The zeroth-order term vanishes at $r=3M$. Expanding around this point to first order,
\begin{equation}
rA_0'(r) - 2A_0(r)
\simeq \left.\frac{d}{dr}\big(rA_0'(r) - 2A_0(r)\big)\right|{r=3M} (r - 3M),
\end{equation}
and substituting $r = 3M + \epsilon \delta r$, we obtain
\begin{equation}
\left.\frac{d}{dr}\big(rA_0'(r) - 2A_0(r)\big)\right|{r=3M} \epsilon \delta r \epsilon \big[3M a'(3M) - 2a(3M)\big] = 0,
  \end{equation}
where terms of order $\mathcal{O}(\epsilon^2)$ and higher have been neglected within the perturbative expansion.
Evaluating the derivative for the Schwarzschild background,
\begin{equation}
\left.\frac{d}{dr}\big(rA_0'(r) - 2A_0(r)\big)\right|_{r=3M}
= -\frac{2}{3M},
\end{equation}
we obtain the first-order correction to the photon-sphere radius,
\begin{equation}
\delta r = \frac{3M}{2}\big[3M a'(3M) - 2a(3M)\big].
\end{equation}
Photon spheres and strong-field black-hole observables in ghost-free $f(G)$ gravity were recently studied by Nojiri and Odintsov~\cite{Nojiri:2024nlx}. Their analysis showed that Gauss--Bonnet corrections can significantly modify the structure of unstable photon orbits and the associated black-hole shadow. These results are closely related to the present work, where the higher-curvature $f(R,\mathcal{G})$ corrections induce shifts in the photon-sphere radius and shadow size.

Using the asymptotic expansion of the perturbation function
$a(r)$ derived from the field equations, one finds that the correction takes the leading-order form
\begin{equation}
\delta r \propto \frac{\beta}{M^5} + \frac{\gamma}{M^7},
\end{equation}
showing that higher-curvature effects modify the photon-sphere radius through the coupling constants $\beta$ and $\gamma$. It is important to emphasize that this result has been obtained using an asymptotic expansion of $a(r)$ valid at large $r$.

\subsection*{Physical Interpretation}

The photon sphere determines the critical impact parameter separating captured and scattered photon trajectories and therefore directly controls the size of the black hole shadow\cite{Synge:1966okc,Perlick:2004tq}. A positive correction $\delta r$ shifts the photon sphere outward, increasing the apparent shadow size, while a negative correction shifts it inward.

Since the correction depends explicitly on the higher-curvature couplings, observations of black-hole shadows provide, in principle, a way to constrain deviations from general relativity in the strong-field regime. Alternative modifications of black-hole geometries have also been proposed in the context of Lorentzian--Euclidean black holes, where the spacetime signature changes across the horizon. In this framework, Capozziello \textit{et al.}~\cite{Capozziello:2024ucm} showed that singularity-free black-hole configurations can emerge while preserving solutions of the vacuum Einstein equations. The resulting geometries exhibit nontrivial causal and optical properties that differ from the classical Schwarzschild spacetime.


To assess parameter sensitivity, it is convenient to introduce the dimensionless couplings
\begin{equation}
\bar{\beta} \equiv \frac{\beta}{M^6}, \qquad
\bar{\gamma} \equiv \frac{\gamma}{M^8}.
\end{equation}
Within the asymptotic perturbative approximation adopted here, the photon-sphere shift can be estimated as
\begin{equation}\label{38}
\frac{\delta r}{M}
\approx
0.17846\,\bar{\beta}
+
1.25015\,\bar{\gamma},
\end{equation}
where the numerical coefficients are obtained numerically from the asymptotic expansion of the perturbation function evaluated at $r = 3M$. It is important to emphasize that the above estimate is obtained by evaluating the asymptotic expansion of the perturbation function $a(r)$ at $r = 3M$, which lies outside the strict large-$r$ regime where the expansion is formally valid. Consequently, the numerical coefficients in Eq.~(\ref{38}) should be regarded as indicative rather than precise. Nevertheless, the expression captures the leading scaling behavior of the correction and provides a useful qualitative comparison between the different higher-curvature contributions. In particular, it suggests that, for comparable normalized couplings, the contribution from the $\gamma$-sector is significantly larger than that from the $\beta$-sector. This hierarchy is expected to be robust, as it reflects the different curvature structures entering the field equations, even though the precise numerical coefficients may be sensitive to higher-order corrections and a more accurate treatment of the near-photon-sphere region.


\begin{figure}[H]
\centering
\subfigure[~Metric function  $A(r)$ compared with the Schwarzschild case]{\label{fig:csh}\includegraphics[scale=0.27]{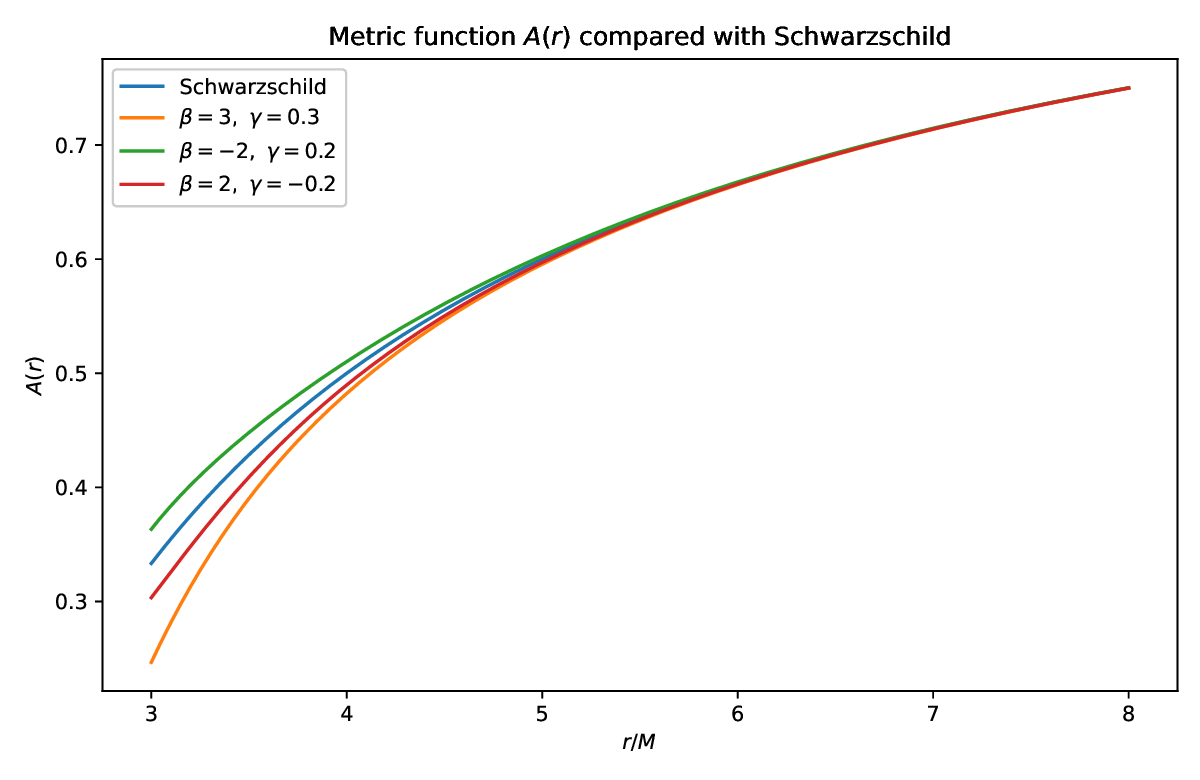}}
\subfigure[~Photon-sphere condition compared with the Schwarzschild case]{\label{fig:psc}\includegraphics[scale=0.27]{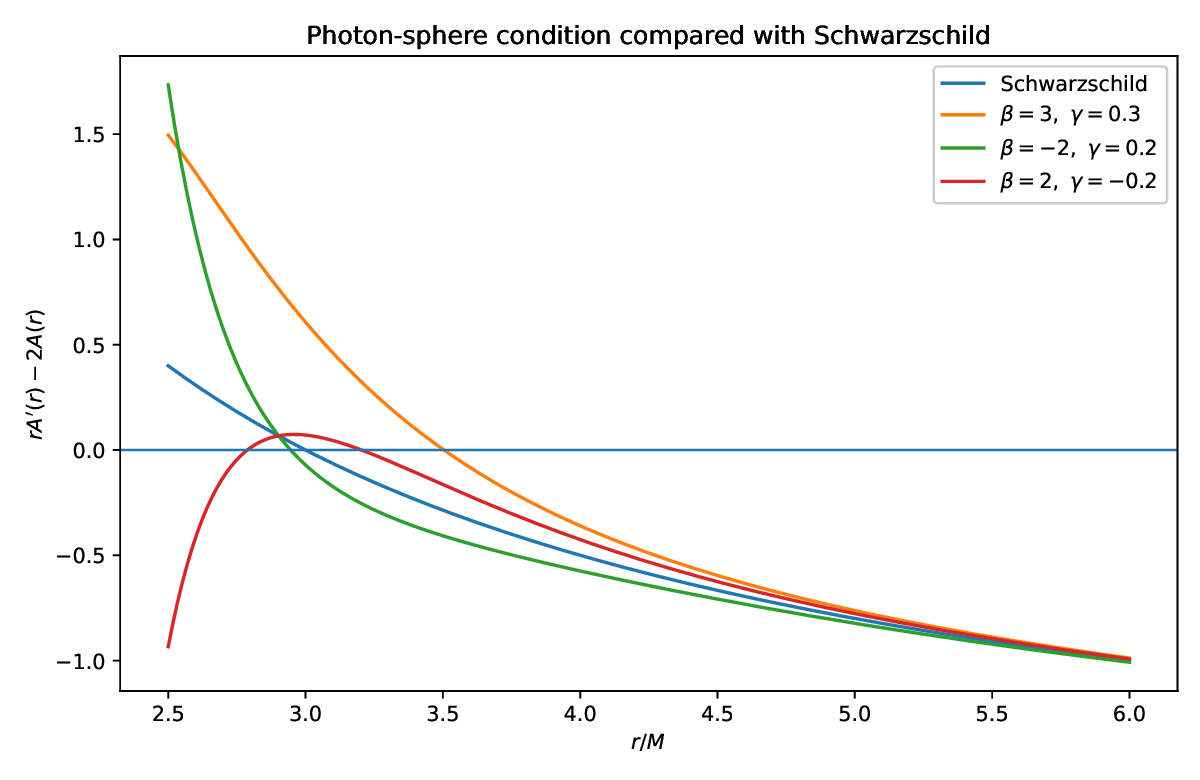}}
\subfigure[~Photon-sphere function compared with the Schwarzschild case]{\label{fig:psf}\includegraphics[scale=0.27]{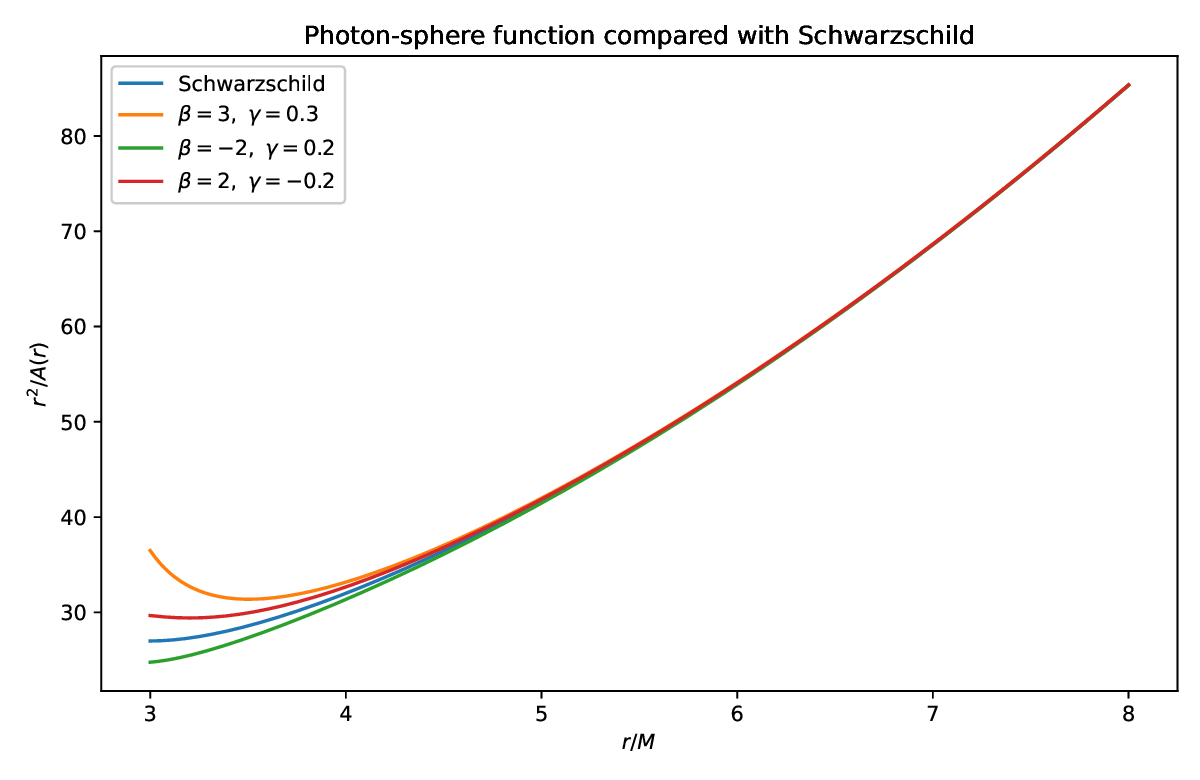}}
\caption{Comparison of the modified metric function, the photon-sphere condition, and the photon-sphere function with their corresponding Schwarzschild behaviors for different values of the coupling parameters.}
\label{Fig:1}
\end{figure}
Figure~\ref{Fig:1} illustrates the impact of higher-curvature corrections on the metric and photon-sphere structure. Panel \subref{fig:csh} shows that the modified metric function $A(r)$ remains close to the Schwarzschild solution, validating the perturbative approach, while exhibiting noticeable deviations near the strong-field region. Panel \subref{fig:psc} demonstrates that the photon-sphere condition is shifted due to the presence of the coupling parameters, indicating a modification of null circular orbits. Panel \subref{fig:psf} further confirms this behavior through the corresponding photon-sphere function, whose variation directly translates into a shift of the photon-sphere radius. Overall, the plots support the analytical results and show that even small higher-curvature corrections can produce measurable effects in the strong-gravity regime.
\section{Black-Hole Shadow}\label{V}
The shadow of a black hole corresponds to the apparent dark region observed by a distant observer, formed by photons that are captured by the black hole rather than reaching the observer. The boundary of the shadow is determined by unstable circular photon orbits, i.e., the photon sphere, and therefore provides an observational window into the structure of spacetime under extreme gravitational conditions \cite{Synge:1966okc,Bardeen:1973tla,Perlick:2004tq}.  The optical appearance of quantum-corrected black holes has recently been explored in effective-field-theory models of gravity. Wang and Battista~\cite{Wang:2025fmz} investigated the properties of unstable photon orbits, black-hole shadows, and photon rings in quantum-corrected Schwarzschild spacetimes, showing that quantum effects can produce observable deviations in the shadow structure and optical image of the black hole.

For a static, spherically symmetric spacetime given by Eq~\eqref{met}
the motion of photons is governed by null geodesics. As shown in the previous section, the photon-sphere radius $r_{\mathrm{ph}}$ is determined by Eq.~\eqref{ps}.


For an observer situated far from the black hole, effectively at spatial infinity, the apparent radius of the shadow is determined by the critical impact parameter \(b_c\), defined as the ratio of the photon's angular momentum \(L\) to its energy \(E\),
\begin{equation}
b_c = \frac{L}{E}.
\end{equation}
Using the null geodesic equations, one finds that the critical impact parameter corresponding to unstable circular photon orbits is given by \cite{Bardeen:1973tla,Perlick:2004tq}
\begin{equation}
b_c^2 = \frac{r_{_{\mathrm{ph}}}^2}{A(r_{_{\mathrm{ph}}})}.
\end{equation}
Thus, the shadow radius observed at infinity is
\begin{equation}
R_{\mathrm{sh}} = b_c = \frac{r_{\mathrm{_{ph}}}}{\sqrt{A(r_{_{\mathrm{ph}}})}}.
\end{equation}

For the Schwarzschild solution, $A_0(r) = 1 - \frac{2M}{r}$, the photon-sphere radius is given by $r_{\mathrm{ph}}^{(0)} = 3M$. Substituting this into the expression for the shadow radius, one obtains the well-known result
\begin{equation}
R_{\mathrm{sh}}^{(0)} = \frac{3M}{\sqrt{1 - \frac{2M}{3M}}}
= 3\sqrt{3}\,M,\qquad \mbox{which defines the shadow size in general relativity \cite{Chandrasekhar:1984siy}.}
\end{equation}

We now consider the corrections to the shadow radius arising from modified gravity within the perturbative framework.  Using the perturbative expansion
\begin{equation}\label{exp}
A(r) = A_0(r) + \epsilon a(r), \qquad
r_{\mathrm{ph}} = 3M + \epsilon \delta r,\quad \mbox{we expand the shadow radius to first order in $\epsilon$.}
\end{equation}
First, expanding the metric function at the photon-sphere radius, we obtain
\begin{equation}
A(r_{\mathrm{ph}}) = A_0(3M) + \epsilon \left[a(3M) + \delta r\, A_0'(3M)\right].
\end{equation}
Since
\begin{equation}
A_0(3M) = \frac{1}{3}, \quad
A_0'(3M) = \frac{2}{9M}, \quad \mbox{this becomes} \quad
A(r_{\mathrm{ph}}) = \frac{1}{3} + \epsilon \left[a(3M) + \frac{2\delta r}{9M}\right].
\end{equation}

Next, expanding the square root,
\begin{equation}
\sqrt{A(r_{\mathrm{ph}})} \simeq \frac{1}{\sqrt{3}} \left[1 + \frac{3\epsilon}{2}\left(a(3M) + \frac{2\,\delta r}{9M}\right)\right].
\end{equation}
Expanding consistently to first order in $\epsilon$, we obtain,
\begin{equation}
R_{\mathrm{sh}} = \frac{3M + \epsilon \delta r}{\sqrt{A(r_{\mathrm{ph}})}},\quad \mbox{and expanding to first order, we obtain} \quad
R_{\mathrm{sh}} = 3\sqrt{3}\,M + \epsilon \delta R.
\end{equation}
where the correction is given by
\begin{equation}
\delta R = \sqrt{3}\left[\delta r - \frac{3M}{2}a(3M) - \frac{\delta r}{3}\right].
\end{equation}

This expression shows that the modification to the shadow radius arises from two distinct contributions:
\begin{itemize}
\item the shift in the photon-sphere radius $\delta r$,
\item the direct correction to the metric function $a(r)$ evaluated at $r=3M$.
\end{itemize}

Using the result obtained in the previous section,
\begin{equation}\label{del}
\delta r \sim \frac{\beta}{M^5} + \frac{\gamma}{M^7}, \quad \mbox{the correction to the shadow radius takes the schematic form} \quad
\delta R \sim \frac{\beta}{M^5} + \frac{\gamma}{M^7}.
\end{equation}

\subsection*{Physical Interpretation}

It is important to emphasize that the present result for the shadow correction is obtained using an asymptotic expansion of the metric perturbation $a(r)$, which is formally valid in the large-$r$ regime. In the present analysis, this expansion is evaluated at $r = 3M$, corresponding to the photon-sphere radius, and therefore lies outside the strict domain of validity of the asymptotic approximation. Consequently, the expression for $\delta R$ should be interpreted as providing a leading-order estimate rather than a quantitatively precise prediction.

Nevertheless, the result captures the dominant scaling behavior of the correction and clearly identifies the two distinct physical contributions: the shift of the photon-sphere radius and the direct modification of the metric function. These features are expected to be robust and to persist in a more complete treatment based on solutions valid in the near-horizon region.



 \begin{figure}[H]
\centering
\includegraphics[width=0.32\linewidth]{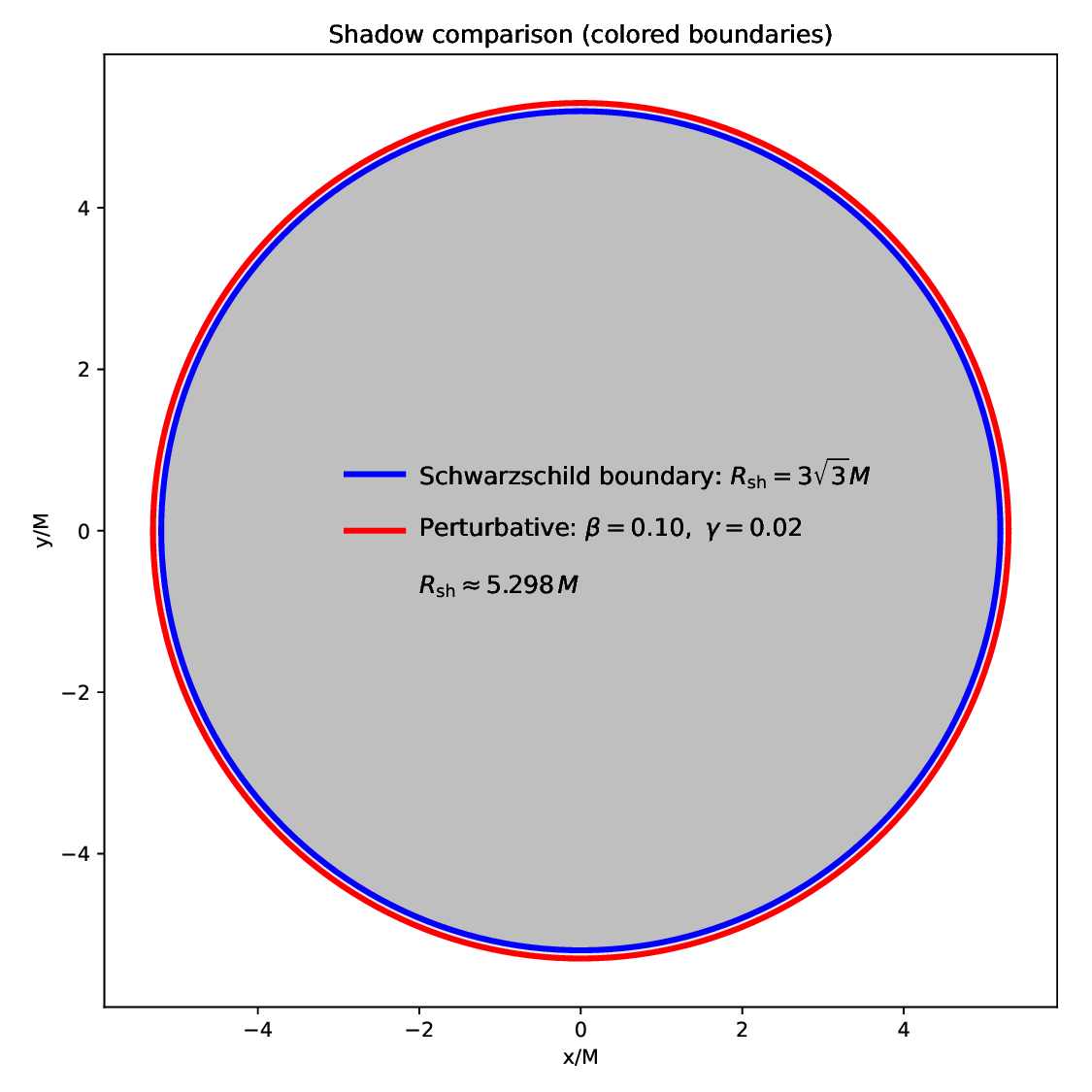}
\caption{Comparison of the black-hole shadow boundary in the modified $f(R,G)$ gravity model with the Schwarzschild case. The deviation from the general relativity prediction is small and consistent with the perturbative framework, while still indicating a measurable shift in the shadow size due to higher-curvature corrections.}
\label{fig:2}
\end{figure}
Figure~\ref{fig:2} illustrates the black-hole shadow boundary in the modified $f(R,G)$ gravity model compared to the Schwarzschild case. The shadow retains its circular shape, as expected for a static and spherically symmetric spacetime, while its radius exhibits a small deviation from the general relativity prediction. This deviation reflects the influence of higher-curvature corrections on the photon-sphere structure and, consequently, on the shadow size. The results are consistent with the perturbative analysis, confirming that the corrections remain well-controlled while still producing observable effects in the strong-field regime.

\section{Strong Gravitational Lensing}\label{VI}

Gravitational lensing provides one of the most powerful observational probes of the spacetime geometry around compact objects. In particular, in the strong-field regime near black holes, light rays can undergo large deflection angles, leading to the formation of relativistic images and characteristic observational signatures \cite{Virbhadra:1999nm,Bozza:2002zj,Perlick:2004tq}. Strong gravitational lensing has also been extensively explored as a probe of extra dimensions and higher-curvature corrections. In particular, Chakraborty and SenGupta~\cite{Chakraborty:2016lxo} showed that strong-deflection observables are highly sensitive to modifications of the spacetime geometry induced by extra-dimensional effects and Kalb--Ramond fields. Their analysis further supports the idea that lensing observables near the photon sphere provide a promising avenue for testing deviations from general relativity in the strong-field regime.

For a static, spherically symmetric spacetime described by the metric in Eq.~\eqref{met}, photon motion is governed by null geodesics. Using the conserved energy $E$ and angular momentum $L$, the deflection angle of a photon coming from infinity and reaching a minimum radial distance $r_0$ is given by \cite{Weinberg:1972kfs}
\begin{equation}
\alpha(r_0) = -\pi + 2 \int_{r_0}^{\infty}
\frac{dr}{\sqrt{B(r)}}
\left[
\frac{1}{\sqrt{\dfrac{r^2}{b^2 A(r)} - 1}}
\right],
\end{equation}
where the impact parameter $b$ is related to $r_0$ through
\begin{equation}
b^2 = \frac{r_0^2}{A(r_0)}.
\end{equation}
The deflection angle is evaluated numerically using the perturbative metric functions obtained in Sec.~\ref{III}. In practice, we truncate the asymptotic expansion of the metric at leading nontrivial order and substitute it into the integral expression. The resulting integral is computed for representative values of the coupling parameters, ensuring that the perturbative conditions remain satisfied.

It is convenient for numerical analysis to express the deflection angle entirely in terms of the closest approach distance $r_0$. Using the relation above, the deflection angle can be rewritten as
\begin{equation}
\alpha(r_0)
=
-\pi
+
2\int_{r_0}^{\infty}
\frac{r_0\,\sqrt{A(r)B(r)}}{\sqrt{r^2A(r_0)-r_0^2A(r)}}\,dr.
\label{alpha_r0_final}
\end{equation}

\begin{figure}[H]
\centering
\subfigure[~Deflection angle $\alpha(r_0)$ versus $r_0/M$, highlighting the divergence near the photon sphere.]{\label{fig:def}\includegraphics[scale=0.3]{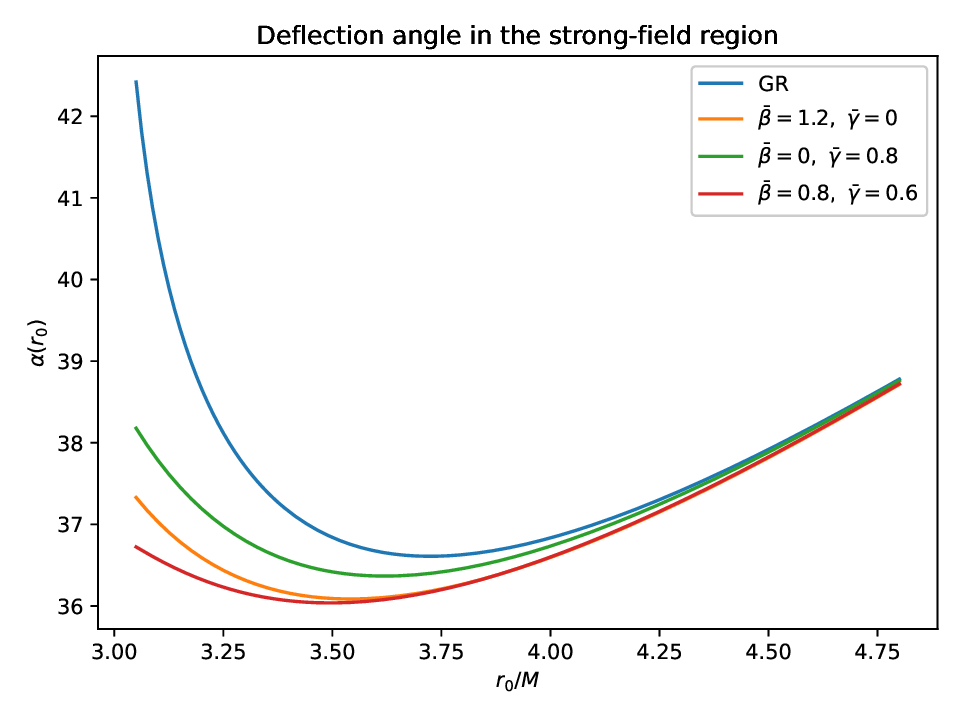}}
\centering
\subfigure[Relative deviation of the deflection angle versus $r_0/M$, highlighting strong-field effects.]{\label{fig:div}\includegraphics[scale=0.3]{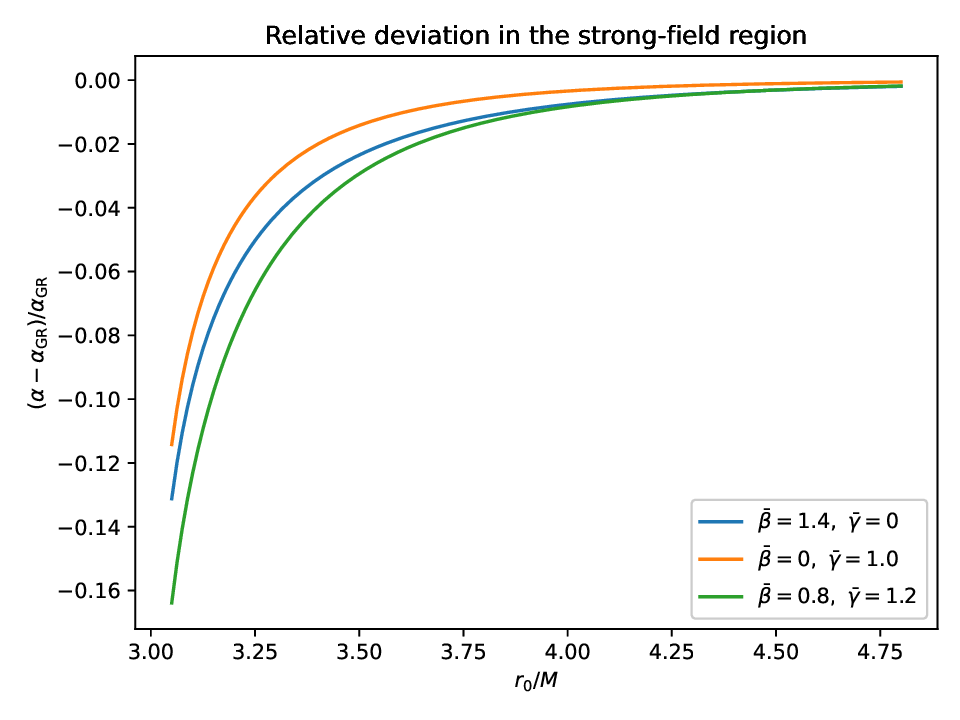}}
\caption{Deflection angle and its deviation from the Schwarzschild case in the strong-field regime. \subref{fig:def} Deflection angle $\alpha(r_0)$ as a function of the closest approach distance $r_0/M$ for the Schwarzschild solution and for representative values of the normalized coupling parameters. The rapid growth as $r_0 \to r_{\mathrm{ph}}$ reflects the expected logarithmic divergence near the photon sphere. \subref{fig:div} Relative deviation $(\alpha-\alpha_{\mathrm{GR}})/\alpha_{\mathrm{GR}}$ as a function of $r_0/M$. The deviation remains small at large distances but becomes more pronounced in the strong-field region, indicating that gravitational lensing is sensitive to higher-curvature corrections.}
\label{Fig:3}
\end{figure}

This form is particularly suitable for numerical evaluation and is the expression used to generate Fig.~\ref{Fig:3}.

\medskip

In the strong-field limit, when $r_0 \to r_{\mathrm{ph}}$, the deflection angle diverges logarithmically.   The behavior of null geodesics and strong-field optical observables has also been investigated in Lorentzian--Euclidean black-hole geometries. Capozziello, Battista, and De Bianchi~\cite{Capozziello:2025wwl} analyzed photon trajectories, causal structure, and matter accretion in such spacetimes, showing that strong gravitational lensing and related optical effects are highly sensitive to deviations from the Schwarzschild geometry. This occurs because the denominator of the integrand in Eq.~\eqref{alpha_r0_final} approaches zero near the photon sphere, producing a logarithmic singularity. This behavior can be captured analytically in the standard form \cite{Bozza:2002zj}
\begin{equation}
\alpha(b) \simeq -\bar{a} \ln\left(\frac{b}{b_c} - 1\right) + \bar{b},
\end{equation}
where $b_c$ is the critical impact parameter corresponding to the photon sphere, and $\bar{a}$ and $\bar{b}$ are model-dependent coefficients determined by the metric functions and their derivatives evaluated at $r_{\mathrm{ph}}$.

In general relativity, one has $r_{\mathrm{ph}} = 3M$ and $b_c = 3\sqrt{3}M$. In modified gravity, corrections to the metric functions shift both the photon-sphere radius and the critical impact parameter.

Within the perturbative $f(R,G)$ framework, the metric functions take the form
\begin{equation}
A(r) = A_0(r) + \epsilon\, a(r), \qquad
B(r) = B_0(r) + \epsilon\, b(r),
\end{equation}
with $A_0(r) = 1 - \frac{2M}{r}$ and $B_0(r) = A_0(r)^{-1}$. The photon-sphere radius is shifted as
\begin{equation}
r_{\mathrm{ph}} = 3M + \epsilon\, \delta r,
\end{equation}
which induces a corresponding correction to the critical impact parameter,
\begin{equation}
b_c = \sqrt{\frac{r_{\mathrm{ph}}^2}{A(r_{\mathrm{ph}})}} = 3\sqrt{3}M + \epsilon\, \delta b.
\end{equation}

To quantify the effect of higher-curvature corrections on light bending, we expand the deflection angle as
\begin{equation}
\alpha(r_0) = \alpha_{\mathrm{GR}}(r_0) + \epsilon\, \delta\alpha(r_0).
\end{equation}
This representation is particularly convenient since it allows a direct numerical evaluation of the deflection angle without requiring the inversion $b(r_0)$.

Substituting the perturbative metric into Eq.~\eqref{alpha_r0_final} and expanding to first order in $\epsilon$, the correction takes the explicit integral form
\begin{equation}
\delta\alpha(r_0)
=
\int_{r_0}^{\infty} dr\, r_0
\left[
\frac{
\dfrac{a(r)}{A_0(r)} + A_0(r)\, b(r)
}{
\sqrt{D_0(r;r_0)}
}
-
\frac{
r^2 a(r_0) - r_0^2 a(r)
}{
\left[D_0(r;r_0)\right]^{3/2}
}
\right], \quad \mbox{where} \quad  D_0(r;r_0) = r^2 A_0(r_0) - r_0^2 A_0(r). \label{delta_alpha_final}
\end{equation}

This expression makes explicit how the higher-curvature corrections $a(r)$ and $b(r)$ modify the deflection angle and is suitable for direct numerical computation.

\medskip

For comparison with the strong-deflection formalism, one may also express the correction in terms of the impact parameter as
\begin{equation}\label{del1}
\delta \alpha(b) =
-\,\delta \bar{a}\,\ln\!\left(\frac{b}{b_c^{\mathrm{GR}}}-1\right)
+ \frac{\bar{a}_{\mathrm{GR}}}{\frac{b}{b_c^{\mathrm{GR}}}-1}
\left(\frac{b}{(b_c^{\mathrm{GR}})^2}\,\delta b_c\right)
+ \delta \bar{b},
\end{equation}
which highlights the role of the photon-sphere shift and the modifications of the strong-lensing coefficients. In Eq.~\eqref{del1} the first term represents the correction to the logarithmic coefficient $\bar{a}$ while the second term arises from the shift in the critical impact parameter $b_c$, which is directly related to the photon-sphere radius and the third term corresponds to the correction to the constant offset $\bar{b}$.

\medskip

To illustrate these effects, Fig.~\ref{Fig:3} shows the deflection angle $\alpha(r_0)$ and its relative deviation from the Schwarzschild case. The results confirm that the deflection angle increases rapidly as $r_0 \to r_{\mathrm{ph}}$, reflecting the expected logarithmic divergence. Moreover, the deviation remains small at large distances but becomes significant near the photon sphere, indicating that strong gravitational lensing is particularly sensitive to higher-curvature corrections.

The present analysis relies on an asymptotic perturbative expansion of the metric functions. While this approach captures the qualitative behavior of strong gravitational lensing and its sensitivity to higher-curvature effects, the quantitative results near the photon sphere should be regarded as approximate. A more precise treatment would require a solution valid in the near-horizon region and a fully consistent numerical integration of the null geodesics.

\section{Implications for Quasinormal Modes and Ringdown}\label{VII}

Quasinormal modes (QNMs) describe the characteristic damped oscillations of black holes under external perturbations and play a central role in gravitational-wave astronomy. These modes depend only on the background spacetime geometry and therefore provide a powerful tool for testing gravitational theories in the strong-field regime \cite{Kokkotas:1999bd,Berti:2009kk,Konoplya:2011qq}.

In the eikonal limit ($\ell \gg 1$), the quasinormal frequencies are determined by the properties of unstable null geodesics \cite{Cardoso:2008bp,Stefanov:2010xz}. Since both the angular velocity and Lyapunov exponent depend on the metric functions at the photon sphere, the perturbative corrections derived in Sec.~\ref{IV} directly induce corresponding shifts in the quasinormal spectrum.

To investigate QNMs in the present framework, we consider the propagation of a test massless scalar field in the black hole background. The dynamics of the scalar field $\Phi$ is governed by the Klein--Gordon equation
\begin{equation}
\Box \Phi = \frac{1}{\sqrt{-g}} \partial_\mu \left( \sqrt{-g}\, g^{\mu\nu} \partial_\nu \Phi \right) = 0.
\end{equation}

By decomposing the scalar field as
\begin{equation}
\Phi(t,r,\theta,\phi) = \sum_{\ell m} \frac{\Psi_{\ell}(r)}{r} Y_{\ell m}(\theta,\phi) e^{-i \omega t},
\end{equation}
one obtains a Schr\"odinger-like wave equation for the radial function $\Psi_{\ell}(r)$,
\begin{equation}
\frac{d^2 \Psi_{\ell}}{dr_*^2} + \left( \omega^2 - V_{\mathrm{eff}}(r) \right) \Psi_{\ell} = 0,
\end{equation}
where the tortoise coordinate $r_*$ is defined by
\begin{equation}
\frac{dr_*}{dr} = \sqrt{\frac{B(r)}{A(r)}}.
\end{equation}

The effective potential takes the form \cite{Chandrasekhar:1984siy}
\begin{equation}
V_{\mathrm{eff}}(r) = A(r)\left[ \frac{\ell(\ell+1)}{r^2} + \frac{A'(r)}{r} \right].
\end{equation}

In the perturbative $f(R,G)$ framework considered in this work, the metric function is given by Eq.~\eqref{met}. Substituting this form into the effective potential yields
\begin{equation}
V_{\mathrm{eff}}(r) = V_{\mathrm{GR}}(r) + \epsilon\, \delta V(r),
\end{equation}
where $V_{\mathrm{GR}}(r)$ corresponds to the Schwarzschild potential and $\delta V(r)$ encodes the corrections induced by higher-curvature terms.

The quasinormal frequencies are determined by imposing physically motivated boundary conditions: purely ingoing waves at the event horizon and purely outgoing waves at spatial infinity \cite{Kokkotas:1999bd}
\begin{equation}
\Psi \sim e^{-i\omega r_*}, \quad \mbox{as} \quad r_* \to -\infty, \qquad
\Psi \sim e^{+i\omega r_*}, \quad \mbox{as} \quad r_* \to +\infty.
\end{equation}

In the eikonal limit ($\ell \gg 1$), there exists a well-established correspondence between quasinormal modes and null geodesics \cite{Cardoso:2008bp,Stefanov:2010xz}. In this regime, the quasinormal frequencies can be expressed as
\begin{equation}
\omega_{\ell n} \simeq \ell\, \Omega_{\mathrm{ph}} - i \left(n + \frac{1}{2}\right) |\lambda_{\mathrm{ph}}|,
\end{equation}
where $\Omega_{\mathrm{ph}}$ is the angular velocity of the unstable photon orbit and $\lambda_{\mathrm{ph}}$ is the associated Lyapunov exponent governing the instability timescale.

Since both $\Omega_{\mathrm{ph}}$ and $\lambda_{\mathrm{ph}}$ depend on the metric functions evaluated at the photon-sphere radius, the corrections obtained in Sec.~\ref{IV} directly translate into shifts in the quasinormal spectrum. In particular, using the perturbative expansion of Eq.~\eqref{exp} one finds that the quasinormal frequencies acquire corrections of the form
\begin{equation}
\omega_{\ell n} = \omega_{\ell n}^{\mathrm{GR}} + \epsilon\, \delta \omega_{\ell n},
\end{equation}
where $\delta \omega_{\ell n}$ depends explicitly on the higher-curvature couplings $\beta$ and $\gamma$.

Physically, a positive correction to the photon-sphere radius leads to a reduction in the oscillation frequency and modifies the damping rate, while a negative correction produces the opposite effect. Therefore, quasinormal modes provide a complementary probe to shadow and lensing observables, allowing one to test modified gravity effects through gravitational-wave measurements.

It is important to emphasize that the present analysis relies on the perturbative expansion of the metric functions and the eikonal approximation. A more accurate determination of the quasinormal spectrum would require solving the full perturbation equations numerically for the modified background \cite{Konoplya:2011qq}. Nevertheless, the present approach captures the leading-order behavior and provides valuable insight into how higher-curvature corrections affect the ringdown signal of black holes.
 \begin{figure}[H]
\centering
\includegraphics[width=0.32\linewidth]{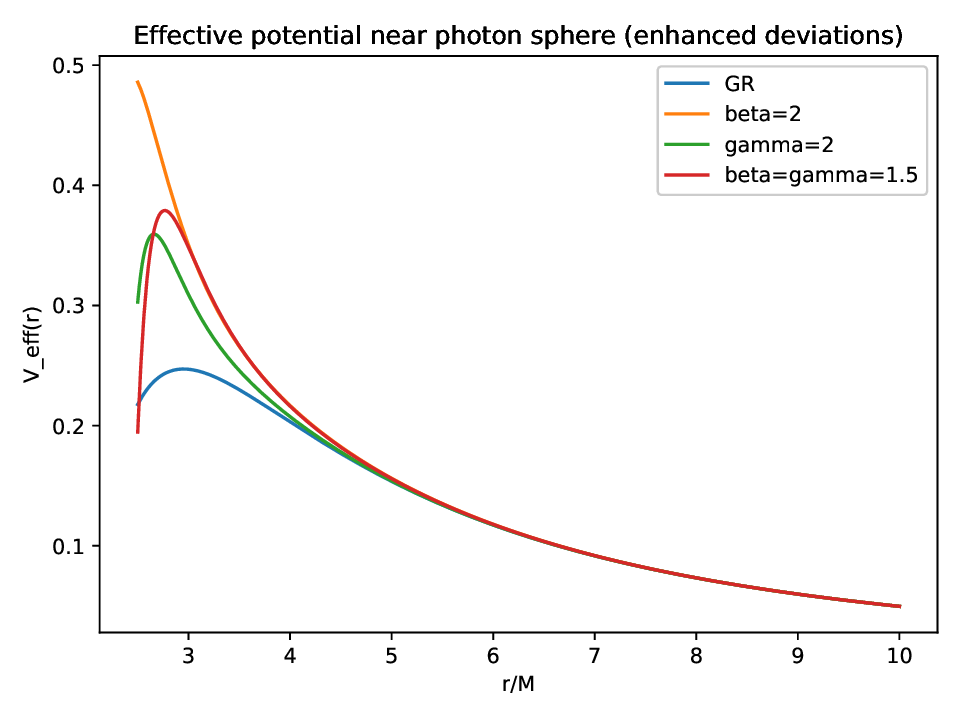}
\caption{
Effective potential \(V_{\mathrm{eff}}(r)\) in the vicinity of the photon sphere for the Schwarzschild spacetime and for several representative choices of the higher-curvature coupling parameters. The plot highlights the enhancement of deviations induced by the $f(R,G)$ corrections in the strong-field region. While the potential coincides with the Schwarzschild case at large distances, noticeable differences arise near the photon sphere, leading to shifts in the peak position and height of the potential. These modifications directly affect the properties of null geodesics and are reflected in observable quantities such as the deflection angle and quasinormal modes.
}
\label{fig:4}
\end{figure}
To further illustrate the impact of higher-curvature corrections on null geodesics, Fig.~\ref{fig:4} shows the behavior of the effective potential near the photon sphere. The results indicate that, while the potential coincides with the Schwarzschild case at large distances, noticeable deviations arise in the strong--field region. In particular, the position and height of the potential peak are shifted due to the coupling parameters, reflecting the modification of unstable photon orbits. Since the quasinormal frequencies are directly related to the properties of the effective potential at the photon sphere, these deviations translate into corrections to the oscillation frequencies and damping rates.

It should be emphasized that the present analysis provides only an approximate description based on the eikonal limit and the perturbative expansion of the metric functions. A complete determination of the quasinormal mode spectrum would require solving the perturbation equations for the modified background, for example using numerical or WKB methods. The results presented here should therefore be interpreted as qualitative estimates of how higher-curvature corrections affect the ringdown signal.

\section{Discussion}\label{VIII}

In this work, we have analyzed the impact of higher-curvature corrections in perturbative $f(R, G)$ gravity on key strong-field observables associated with black holes. By constructing analytic deviations from the Schwarzschild geometry, we have shown that even small corrections to the gravitational action can produce measurable effects on null geodesics and related observables.

A central result of our analysis is the shift in the photon-sphere radius induced by the higher-curvature couplings. Since the photon sphere governs both gravitational lensing and the black-hole shadow, its modification provides a direct link between the underlying theory of gravity and observable quantities \cite{Claudel:2000yi,Perlick:2004tq}. We find that the contribution from the Gauss--Bonnet sector is typically more pronounced than that from mixed curvature terms, indicating that different invariants leave distinguishable imprints in the strong-field regime.

The resulting corrections to the shadow radius remain perturbatively small but potentially observable. Given the recent horizon-scale imaging of supermassive black holes by the Event Horizon Telescope \cite{EventHorizonTelescope:2019dse,EventHorizonTelescope:2022wkp}, such deviations offer a promising avenue for testing extensions of general relativity. In particular, precise measurements of the shadow size and shape may constrain higher-curvature couplings and discriminate between competing modified gravity models \cite{EventHorizonTelescope:2020qrl,Broderick:2013rlq}.

Our results for strong gravitational lensing further emphasize the enhanced sensitivity of light propagation near the photon sphere. As expected, deviations from general relativity become significant in the strong-deflection regime, where the deflection angle exhibits a logarithmic divergence \cite{Bozza:2002zj}. This behavior suggests that relativistic images and lensing observables could serve as complementary probes of higher-curvature effects.

In addition, we have shown that modifications to the photon-sphere structure directly affect the quasinormal mode spectrum through the well-known correspondence between null geodesics and black-hole ringdown in the eikonal limit \cite{Cardoso:2008bp,Stefanov:2010xz}.  Quantum and semiclassical aspects of modified gravity have also been explored in the framework of $f(R,T)$ theories. In particular, the role of spin effects in particle creation and black-hole evaporation was recently analyzed in Ref.~\cite{AraujoFilho:2025zzf}, where it was shown that modified gravity corrections can influence evaporation mechanisms and quantum particle production processes. These results highlight the broader impact of modified gravitational dynamics on both classical and quantum aspects of black-hole physics. Consequently, gravitational-wave observations provide an independent and complementary channel for testing deviations from general relativity in the strong-field regime \cite{Berti:2009kk}.

It is important to note that our analysis relies on a perturbative expansion valid for small coupling parameters, as well as on asymptotic expressions for the metric perturbations. While this approach captures the leading-order corrections and provides analytical insight, it may not fully describe the near-horizon region. A more complete treatment would require solving the modified field equations beyond the asymptotic regime and extending the analysis to higher orders.

Future work may generalize the present framework to rotating black holes, where deviations from general relativity are expected to produce richer phenomenology, including distortions of the shadow shape and modifications to frame-dragging effects. Such extensions, together with increasingly precise observational data, will further enhance the role of black-hole physics as a probe of fundamental gravitational theories. Alternative nonclassical black-hole geometries have also been proposed in the context of Lorentzian--Euclidean spacetimes. In particular, Capozziello \textit{et al.}~\cite{Capozziello:2025wwl} showed that singularity-free black-hole configurations may emerge through signature-changing geometries, providing a novel framework in which strong-field gravitational observables can deviate from the predictions of classical general relativity. Black-hole shadows and photon-sphere observables have also been investigated in modified mimetic gravity models with nontrivial scalar potentials. In particular, Nojiri and Odintsov~\cite{Nojiri:2024txy} showed that scalar-field modifications of the gravitational sector can alter the structure of black-hole geometries and produce observable deviations in photon trajectories and shadow characteristics. Such studies further emphasize the importance of strong-field optical observables as probes of alternative theories of gravity.
\appendix
\section{Derivation of the Perturbation Equations}

We introduce a bookkeeping parameter $\epsilon$ and expand the metric functions as
\begin{equation}
A(r) = 1 - \frac{2M}{r} + \epsilon\, a(r),
\qquad
B(r) = \left(1 - \frac{2M}{r}\right)^{-1} + \epsilon\, b(r).
\end{equation}
Substituting this expansion into the field equations \eqref{fe} and expanding order by order in $\epsilon$, we obtain:
\paragraph{Zeroth order: $\mathcal{O}(\epsilon^0)$}
At leading order, one recovers the Einstein equations,
\begin{equation}
R_{\mu\nu}^{(0)} - \frac{1}{2} g_{\mu\nu}^{(0)} R^{(0)} = 0,\quad \mbox{whose solution is the Schwarzschild metric}\quad
A_0(r) =\frac{1}{B_0(r)}= 1 - \frac{2M}{r}.
\end{equation}

\paragraph{First order: $\mathcal{O}(\epsilon)$}

At linear order, the field equations reduce to a system of linear differential equations for the perturbations $a(r)$ and $b(r)$, where the metric is decomposed as $g_{\mu\nu} = g_{\mu\nu}^{(0)} + h_{\mu\nu}$, with $h_{\mu\nu}$ representing the metric perturbations:
\begin{equation}
\delta R_{\mu\nu}
- \frac{1}{2} g_{\mu\nu}^{(0)} \delta R
- \frac{1}{2} h_{\mu\nu} R^{(0)}
+ (g_{\mu\nu}^{(0)} \Box - \nabla_\mu \nabla_\nu)\,\delta f_R
+ \delta \mathcal{H}_{\mu\nu}
= 0.
\end{equation}
For a static, spherically symmetric spacetime, we consider the line element \eqref{met} and expand the metric functions perturbatively as given in Eq.~\eqref{prt}
where $A_0(r) = 1 - \frac{2M}{r}$ and $B_0(r) = \left(1 - \frac{2M}{r}\right)^{-1}$ correspond to the Schwarzschild solution.

Substituting this expansion into the field equations and keeping only terms up to $\mathcal{O}(\epsilon)$, the modified field equations reduce to a system of linear ordinary differential equations for the perturbations $a(r)$ and $b(r)$.

In particular, the $(tt)$ and $(rr)$ components of the field equations yield
\begin{align}
a'(r) + \frac{a(r)}{r} + \frac{b(r)}{r\left(1-\frac{2M}{r}\right)}
&= \mathcal{S}_1(r), \label{eq:a_eq} \\
b'(r) - \frac{b(r)}{r} - \frac{a(r)}{r\left(1-\frac{2M}{r}\right)}
&= \mathcal{S}_2(r), \label{eq:b_eq}
\end{align}
where the left-hand sides arise purely from the linearized Einstein tensor, while the right-hand sides encode the deviations from general relativity due to higher-curvature corrections.

The source terms $\mathcal{S}_1(r)$ and $\mathcal{S}_2(r)$ originate from the perturbations of $f_R$ and $f_{\mathcal{G}}$ and from the Gauss--Bonnet contribution $\mathcal{H}_{\mu\nu}$. Schematically, they can be written as
\begin{equation}
\mathcal{S}_i(r) = \mathcal{S}_i^{(R^2)}(r) + \mathcal{S}_i^{(R\mathcal{G})}(r) + \mathcal{S}_i^{(\mathcal{G}^2)}(r),
\end{equation}
with contributions given by
\begin{align}
\mathcal{S}_i^{(R^2)}(r) &\sim \alpha \left[ \Box R^{(0)} + \nabla\nabla R^{(0)} \right], \qquad
\mathcal{S}_i^{(R\mathcal{G})}(r) &\sim \beta \left[ \mathcal{G}^{(0)} R^{(0)} + \nabla\nabla \mathcal{G}^{(0)} \right], \qquad
\mathcal{S}_i^{(\mathcal{G}^2)}(r) &\sim \gamma \left[ \mathcal{G}^{(0)} \Box \mathcal{G}^{(0)} + (\nabla \mathcal{G}^{(0)})^2 \right].
\end{align}

More explicitly, using $f_R = 1 + 2\alpha R + \beta \mathcal{G}$ and $f_{\mathcal{G}} = \beta R + 2\gamma \mathcal{G}$, the source terms take the form
\begin{equation}
\mathcal{S}_i(r) \propto
\left( g_{\mu\nu}^{(0)} \Box - \nabla_\mu \nabla_\nu \right)
\left( 2\alpha R^{(0)} + \beta \mathcal{G}^{(0)} \right)
+ \delta \mathcal{H}_{\mu\nu}\big[ R^{(0)}, \mathcal{G}^{(0)} \big], \quad \mbox{evaluated on the Schwarzschild background.}
\end{equation}
For the model under consideration  given by Eq.~\eqref{mod} we have the derivatives of $f_R$ and $f_{\mathcal G}$ give by Eq.~\eqref{diff}. On the Schwarzschild background one has
\begin{equation}
R^{(0)}_{\mu\nu}=0,
\qquad
R^{(0)}=0,
\qquad
A_0(r)=1-\frac{2M}{r},
\end{equation}
so that
\begin{equation}
f_R^{(0)}=1+\beta \mathcal G^{(0)},
\qquad
f_{\mathcal G}^{(0)}=2\gamma \mathcal G^{(0)}.
\end{equation}
Moreover, the Gauss--Bonnet invariant is
\begin{equation}
\mathcal G^{(0)}=\frac{48M^2}{r^6}.
\end{equation}

Therefore, at first order the higher-curvature sector enters through the effective source tensor
\begin{equation}
\mathcal J_{\mu\nu}^{(0)}
=
\beta\Bigl(g_{\mu\nu}^{(0)}\Box-\nabla_\mu\nabla_\nu\Bigr)\mathcal G^{(0)}
+8\gamma\,R^{(0)}_{\mu\lambda\nu\sigma}\nabla^\lambda\nabla^\sigma \mathcal G^{(0)}
-\gamma\,g_{\mu\nu}^{(0)}\bigl(\mathcal G^{(0)}\bigr)^2 .
\label{eq:Jmunu-def}
\end{equation}
Thus the linearized field equations may be written schematically as
\begin{equation}
\delta G_{\mu\nu}+\mathcal J_{\mu\nu}^{(0)}=0.
\end{equation}

Since $\mathcal G^{(0)}$ depends only on $r$, its derivatives are
\begin{align}
\mathcal G^{(0)\prime}(r)&=-\frac{288M^2}{r^7},\qquad
\mathcal G^{(0)\prime\prime}(r)=\frac{2016M^2}{r^8}.
\end{align}
For a scalar $\Phi(r)$ in the Schwarzschild geometry,
\begin{equation}
\Box \Phi
=
A_0\,\Phi''+\left(\frac{2A_0}{r}+A_0'\right)\Phi',\quad \mbox{hence}\qquad
\Box \mathcal G^{(0)}
=
\frac{1440M^2}{r^8}
-\frac{3456M^3}{r^9}.
\end{equation}
The nonvanishing second covariant derivatives are
\begin{align}
\nabla_t\nabla_t \mathcal G^{(0)}
&=
-\Gamma^{r}_{tt}\,\mathcal G^{(0)\prime}
=
\frac{288A_0 M^3}{r^9},\qquad
\nabla_r\nabla_r \mathcal G^{(0)}
=
\mathcal G^{(0)\prime\prime}
+\frac{A_0'}{2A_0}\mathcal G^{(0)\prime}
=
\frac{2016M^2}{r^8}
-\frac{288M^3}{A_0 r^9},\\
\nabla_\theta\nabla_\theta \mathcal G^{(0)}
&=
-\Gamma^r_{\theta\theta}\,\mathcal G^{(0)\prime}
=
-\frac{288A_0M^2}{r^6},\quad \mbox{with} \qquad
\nabla_\phi\nabla_\phi \mathcal G^{(0)}
=
\sin^2\theta\,\nabla_\theta\nabla_\theta \mathcal G^{(0)}.
\end{align}

The Schwarzschild Riemann tensor components needed in \eqref{eq:Jmunu-def} are
\begin{align}
R^{(0)}_{trtr}&=-\frac{2M}{r^3},\qquad
R^{(0)}_{t\theta t\theta}=\frac{M(r-2M)}{r^2},\qquad
R^{(0)}_{r\theta r\theta}=-\frac{M}{r-2M}.
\end{align}
Using these expressions, the mixed components of the effective source tensor become
\begin{align}
\mathcal J^{t}{}_{t}
&=
\beta\left(\Box \mathcal G^{(0)}-\nabla^{t}\nabla_{t}\mathcal G^{(0)}\right)
+8\gamma\,R^{(0)t}{}_{\lambda t\sigma}\nabla^\lambda\nabla^\sigma \mathcal G^{(0)}
-\gamma\bigl(\mathcal G^{(0)}\bigr)^2 =
\frac{288\beta M^2}{r^9}\,(5r-11M)
-\frac{2304\gamma M^3}{r^{12}}\,(35M-16r),
\label{eq:Jtt-explicit}
\\[1ex]
\mathcal J^{r}{}_{r}
&=
\beta\left(\Box \mathcal G^{(0)}-\nabla^{r}\nabla_{r}\mathcal G^{(0)}\right)
+8\gamma\,R^{(0)r}{}_{\lambda r\sigma}\nabla^\lambda\nabla^\sigma \mathcal G^{(0)}
-\gamma\bigl(\mathcal G^{(0)}\bigr)^2=
-\frac{576\beta M^2}{r^9}\,(2r-3M)
+\frac{2304\gamma M^3}{r^{12}}\,(2r-7M).
\label{eq:Jrr-explicit}
\end{align}

Hence, the source functions that appear in the radial perturbation equations are derived from these effective components after rewriting the linearized Einstein equations in terms of the metric corrections $a(r)$ and $b(r)$. In other words, when the first-order equations are expressed as in Eqs.~\eqref{eq:a_eq} and \eqref{eq:b_eq}, the functions $\mathcal S_1(r)$ and $\mathcal S_2(r)$ correspond to algebraic combinations of $\mathcal J^{t}{}_{t}$ and $\mathcal J^{r}{}_{r}$, respectively. In the standard perturbative parametrization, they can be schematically identified as
\begin{equation}
\mathcal S_1(r)\propto \mathcal J^{t}{}_{t},
\qquad
\mathcal S_2(r)\propto \mathcal J^{r}{}_{r},
\end{equation}
so that their radial dependence is explicitly
\begin{align}
\mathcal S_1(r)&\sim
\frac{288\beta M^2}{r^9}\,(5r-11M)
-\frac{2304\gamma M^3}{r^{12}}\,(35M-16r),
\\
\mathcal S_2(r)&\sim
-\frac{576\beta M^2}{r^9}\,(2r-3M)
+\frac{2304\gamma M^3}{r^{12}}\,(2r-7M).
\end{align}
Therefore, the source terms decay as inverse powers of $r$, with the $\beta$-sector contributing at order $r^{-8}$--$r^{-9}$ and the $\gamma$-sector at order $r^{-11}$--$r^{-12}$.

Starting from the linearized field equations on the Schwarzschild background, the metric perturbations $a(r)$ and $b(r)$ are determined by the $(t,t)$ and $(r,r)$ components of the modified equations, which reduce to the coupled first-order system given in Eqs.~(A4) and (A5). The higher-curvature corrections enter these equations through the effective source tensor $J^\mu{}_\nu$, whose explicit components $J^t{}_t$ and $J^r{}_r$ are given in Eqs.~(A19) and (A20). In the perturbative treatment adopted here, the source functions $S_1(r)$ and $S_2(r)$ are identified, up to the chosen normalization convention, with the radial dependence of $J^t{}_t$ and $J^r{}_r$, respectively. Substituting these source terms into the coupled differential equations and imposing asymptotic flatness, $a(r)\to0$ and $b(r)\to0$ as $r\to\infty$, one solves the system order by order in inverse powers of $r$. In this way, the leading large-$r$ behavior of $a(r)$ and $b(r)$ is obtained directly from the falloff structure of $J^t{}_t$ and $J^r{}_r$, with the $\beta$-sector generating terms beginning at order $r^{-7}$ and the $\gamma$-sector contributing at order $r^{-10}$. Thus, the functions $a(r)$ and $b(r)$ appearing in the perturbed metric are not guessed independently, but are derived systematically by integrating the linearized field equations sourced by the effective components of the higher-curvature tensor.


\end{document}